\documentclass[aps,prb,showpacs,twocolumn,superscriptaddress]{revtex4-1}
\usepackage{graphicx,amsmath,amssymb}
\usepackage[usenames]{color}
\usepackage{indentfirst}
\usepackage{float}
\usepackage[T1]{fontenc}
\usepackage[colorlinks=true, citecolor=blue, urlcolor=blue, linkcolor=blue ]{hyperref}
\hypersetup{breaklinks=true}
\begin{document}

\bibliographystyle{apsrev4-1}

\title{Parity-dependent phase diagrams in spin-cluster two-leg ladders}
\author{Zongsheng Zhou}
\affiliation{Center for Interdisciplinary Studies $\&$ Key Laboratory for Magnetism and
Magnetic Materials of the MoE, Lanzhou University, Lanzhou 730000, China}
\author{Fuzhou Chen}
\affiliation{Center for Interdisciplinary Studies $\&$ Key Laboratory for Magnetism and
Magnetic Materials of the MoE, Lanzhou University, Lanzhou 730000, China}
\author{Yin Zhong}
\affiliation{Center for Interdisciplinary Studies $\&$ Key Laboratory for Magnetism and
Magnetic Materials of the MoE, Lanzhou University, Lanzhou 730000, China}
\author{Hong-Gang Luo}
\affiliation{Center for Interdisciplinary Studies $\&$ Key Laboratory for Magnetism and
Magnetic Materials of the MoE, Lanzhou University, Lanzhou 730000, China}
\affiliation{Beijing Computational Science Research Center, Beijing 100084, China}
\author{Jize Zhao}
\email{zhaojz@lzu.edu.cn}
\affiliation{Center for Interdisciplinary Studies $\&$ Key Laboratory for Magnetism and
Magnetic Materials of the MoE, Lanzhou University, Lanzhou 730000, China}
\begin{abstract}
Motivated by the recent experiment on $\rm{K_2Cu_3O\left(SO_4\right)_3}$, an edge-shared tetrahedral spin-cluster compound 
[M. Fujihala \textit{et al.}, Phys. Rev. Lett. \textbf{120}, 077201 (2018)], we investigate two-leg spin-cluster ladders with 
the plaquette number $n_p$ in each cluster up to six by the density-matrix renormalization group method. We find that 
the phase diagrams of such ladders strongly depend on the parity of $n_p$. For even $n_p$, the phase diagrams have two phases, 
one is the Haldane phase, and the other is the cluster rung-singlet phase. For odd $n_p$, there are four phases, 
which are a cluster-singlet phase, a cluster rung-singlet phase, a Haldane phase and an even Haldane phase. Moreover, in the latter case 
the region of the Haldane phase increases while that of the cluster-singlet phase and the even Haldane phase shrinks as $n_p$ increases. 
We thus conjecture that in the large $n_p$ limit, the phase diagrams will become independent of $n_p$. By analysing the ground-state energy 
and entanglement entropy we obtain the order of the phase transitions. In particular, for $n_p=1$ there is no 
phase transition between the even Haldane phase and the cluster rung-singlet phase while for other odd $n_p$ there is a first-order phase transition.
Our work provides comprehensive phase diagrams for these cluster-based models and may be
helpful to understand experiments on related materials.
\end{abstract}

\maketitle

\section{Introduction}
An integer is either even or odd, which is known as the parity. The properties of some physical systems associated with different parity
may be fundamentally different\cite{J.Chem.Phys.63.2533(1975),PhysRevB.67.100402,PhysRevB.87.144409,PhysRevB.92.161105,PhysRevB.98.085104}. 
A well-known example stems from Haldane's conjecture\cite{PhysRevLett.50.1153,PhysLettA.93.464} that the lowest excitation 
of spin chains with integer spin is gapful while those with half integer spin is gapless. Correspondingly, the lowest excitation of 
spin-$\frac{1}{2}$ ladders with even legs are gapful  but those with odd legs are gapless\cite{Dagotto618,PhysRevB.54.1009,PhysRevLett.77.1865}. This conjecture was soon confirmed 
by various numerical and experimental works\cite{PhysRevB.33.659, PhysRevLett.56.371, doi:10.1063/1.340736,PhysRevB.45.9798,PhysRevB.53.52} and therefore the gapful phase 
is called Haldane phase. Recently, further theoretical works show that the Haldane phase in spin chains with odd integer  
spins and even integer spins are actually different\cite{PhysRevB.85.075125}. 
The former one is protected by some symmetries, such as time-reversal, spacial inversion and dihedral symmetry but the latter is not although edge states may exist in both of them. 
Hereafter, following literature, we will just call the former one as Haldane $\left(\rm HP\right)$ phase but the latter one as even 
Haldane $\left(\rm EHP\right)$ phase. Now we know that the HP phase is actually a symmetry-protected topological phase\cite{PhysRevB.80.155131,PhysRevB.81.064439, PhysRevB.84.165139, PhysRevB.87.155114}. These theoretical 
progresses have stimulated extensive efforts to search for such topologically 
nontrivial phase in quasi one-dimensional materials as well as artificial structures and the HP phase
has been reported in a variety of experiments\cite{PhysRevLett.93.036401, PhysRevX.5.021026, PhysRevLett.119.185701, PhysRevB.96.155133, PhysRevLett.120.085301}. 

Very recently, evidences for the HP phase in spin-cluster materials were first reported 
in the compound $\rm{K_2Cu_3O\left(SO_4\right)_3}$ by M. Fujihala \textit{et al.}\cite{PhysRevLett.120.077201}. 
This compound consists of edge-shared tetrahedral spin clusters. 
Spin-$\frac{1}{2}$ $\rm{Cu^{2+}}$ sits at the corners of the tetrahedra. These clusters are connected via $\rm{SO_4^{2+}}$ 
along the $b$ axis. In other directions, they are connected via nonmagnetic ions or no exchange path is allowed and thus interactions can be neglected. 
These identify $\rm{K_2Cu_3O\left(SO_4\right)_3}$ as a quasi one-dimensional compound. 
Various experimental measurements in combination with theoretical analysis\cite{PhysRevLett.120.077201,PhysRevB.98.180410} 
reveal that its ground state is an HP phase.   
In addition to $\rm{K_2Cu_3O\left(SO_4\right)_3}$, some other cluster-type one-dimensional materials have also been reported\cite{Chem.Matter.12,PhysRevB.79.024416,PhysRevB.86.180405,PhysRevB.87.144425,PhysRevB.90.184402,PhysRevB.96.214424}, 
such as $\rm{Cu_2Te_2O_5X_2}$  with $\rm{X=Cl, Br}$. These experiments call for a systematical investigation on the low-energy properties of spin-cluster ladders.  

\begin{figure}[t]
\includegraphics[width=1.0\columnwidth]{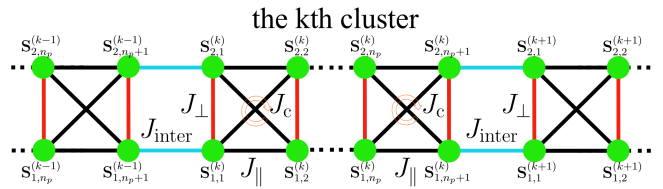}
\caption{The sketch of the cluster-based spin ladder. The interactions corresponding to the model are marked. 
	We have assumed that $J_c=J_\parallel$ therefore the same color represents the same interaction strength. 
	$n_p$ is the number of plaquettes in a single cluster and $k$ is the index of the cluster.
	${\bf{S}}_{i,j}^{(k)}$ represents the spin operators in cluster $k$ with the leg index $i$ and rung index $j$, thus $1\le j \le n_p+1$.}
\label{FIG1}
\end{figure}

For this purpose, we study a Hamiltonian written as follows
\begin{equation}
	\mathcal{H} = \sum_{k=1}^{L_c}\mathcal{H}_{\rm{intra}}^{(k)} + \sum_{k=1} ^{L_c-1} \mathcal{H}_{\rm{inter}}^{(k:k+1)} 
	\label{HAM}
\end{equation}
where $L_c$ is the number of clusters. The Hamiltonian has two parts, the first is the interaction within one cluster, and the other is 
the interaction between two nearest-neighbor clusters. Such intra-cluster and inter-cluster Hamiltonians are given by\cite{PhysRevLett.120.077201}:
\begin{align}
	\mathcal{H}_{\rm{intra}}^{(k)} & = J_\perp\sum_{j=1}^{n_p+1}{\bf{S}}^{(k)}_{1,j}\cdot {\bf{S}}_{2,j}^{(k)}+J_\parallel\sum_{i=1,2}\sum_{j=1}^{n_p}{\bf{S}}_{i,j}^{(k)}\cdot {\bf{S}}_{i,j                            +1}^{(k)}\nonumber\\
		& + J_c\sum_{j=1}^{n_p}\sum_{a=0,1}{\bf{S}}_{1,j+a}^{(k)}\cdot {\bf{S}}_{2,j+1-a}^{(k)}
		\label{HINTRA}
\end{align}	

\begin{align}
	\mathcal{H}_{\rm{inter}}^{(k:k+1)} = J_{\rm{inter}} \sum _{i=1,2}{\bf{S}}_{i,n_p+1}^{(k)}\cdot {\bf{S}}_{i,1}^{(k+1)}
	\label{HINTER}
\end{align}
where $n_p$ is the number of the plaquettes 
within one cluster, which corresponds to the number of the tetrahedra within one cluster in compounds. 
${\bf{S}}_{i,j}^{(k)}$ is the spin operator in the $k$th cluster with the leg index $i$ and rung index $j$.
A schematic representation of the model and the couplings $J_\perp,J_\parallel,J_c$ and $J_{\rm{inter}}$ are plotted in Fig. \ref{FIG1}. 
This model was proposed\cite{PhysRevLett.120.077201} for $\rm{K_2Cu_3O\left(SO_4\right)_3}$, where $n_p$ takes 2 
and $J_c = J_\parallel$ due to the symmetry of a tetrahedron. Although in known compounds, 
$n_p$ is limited to 1 or 2, in our theoretical work we will consider general $n_p$ and extrapolate our conclusions to the 
large $n_p$ limit. Moreover, for simplicity, we assume that $J_c=J_\parallel$ is satisfied for all $n_p$ and set $J_{\rm{inter}}=1$ as the energy unit. 

To study the low-energy properties of Hamiltonian $\left(\ref{HAM}\right)$, we resort to the state-of-art numerical algorithm, 
density-matrix renormalization group (DMRG)\cite{PhysRevLett.69.2863,PhysRevB.48.10345,RevModPhys.77.259,SCHOLLWOCK201196}.
To avoid edge effect, periodic boundary condition (PBC) is employed unless stated explicitly otherwise. We keep up to 3000 optimal bases 
thus the largest truncation error is smaller than $10^{-10}$. 
$U(1)$ symmetry is used to accelerate the computation therefore the Hamiltonian is diagonalized in the sector with fixed $z$-component of the 
total spin. All target states in the given sector are used with equal weight to construct the reduced density matrix.   
Several relevant quantities such as the ground state, the first excited state and corresponding energy $E_0$, $E_1$, 
the entanglement entropy and entanglement spectrum of the ground state are calculated.
All the data in our figures are obtained in the sector with the $z$-component of the total spin zero. As we will show, this is 
enough for our model since all phases are gapful. And then the excitation gap $\Delta\equiv E_1-E_0$. 
To calculate the entanglement entropy and entanglement spectrum, we arrange the ladder into one chain in the rung-major order, 
and split the chain into two halves. One is the system and the other is the environment. 
After tracing out the freedom of the environment, we obtain the reduced density matrix $\rho$. 
The entanglement entropy\cite{RevModPhys.82.277,RevModPhys.90.035007} $S$ is then calculated by its definition 
$S=-\sum_i\rho_i\ln\rho_i$ with $\rho_i$ the eigenvalues of $\rho$. The entanglement spectrum $\xi_i=-\ln\rho_i$ is thus readily available. 

We find that the phase diagrams of this model depend 
strongly on the parity of $n_p$. For even $n_p$, we have two phases, and for odd $n_p$, we have four phases. The particular features for $n_p=1$
and large $n_p$ are also discussed. The rest of the paper is organized as follows.
In Sec. II, we present the phase diagrams for even $n_p$. In Sec. III, we present the phase diagrams for odd $n_p$. In Sec. IV, we show the results in the large $n_p$ limit and in Sec. V we conclude our work.

\section{Phase Diagram for Even $n_p$}
\begin{figure}[t]
\includegraphics[width=1.0\columnwidth, clip]{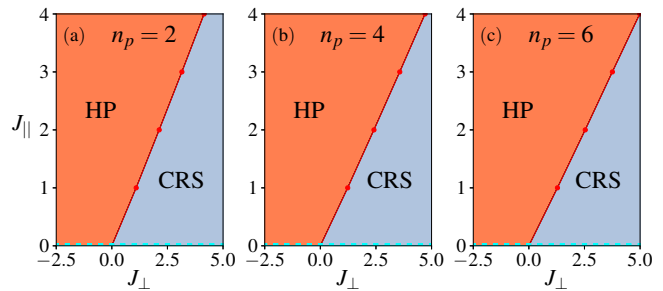}
\caption{Phase diagrams of Hamiltonian (\ref{HAM}) for $n_p = 2,4,6$. There are an HP phase and a CRS phase in each phase diagram(see text for details). 
The transition between the two phases is of the first order. The red solid line is the eye guide of the phase boundary. It's obvious 
that the phase boundary depends almost linearly on $J_\perp$ and as $n_p$ increases the slope of the 
phase boundary decreases. Along the dashed cyan line ($J_\parallel=0$) the rungs within a cluster are decoupled.} 
\label{FIG2}
\end{figure}
In this section, we will discuss the phase diagrams for even $n_p$. In Fig. \ref{FIG2}, we show our results for $n_p=2, 4$ and $6$.  
We find two phases, which are an HP phase and a cluster rung-singlet (CRS) phase. The former is common 
in spin-1 chains and spin-$\frac{1}{2}$ two-leg ladders\cite{PhysRevLett.67.1614,PhysRevLett.56.371,PhysRevLett.80.2713}. 
The latter is a trivial product state of singlets. In this phase, $J_\perp$ dominates over $J_\parallel$, 
and each rung within a cluster (rung index $1<j\le n_p$)
forms a singlet. The four spins in the plaquette connecting the two nearest-neighbor clusters also form a singlet.  
In Appendix \ref{appA}, we provide some numerical evidences for the CRS phase.
\begin{figure}
\includegraphics[width = 1.0\columnwidth, clip]{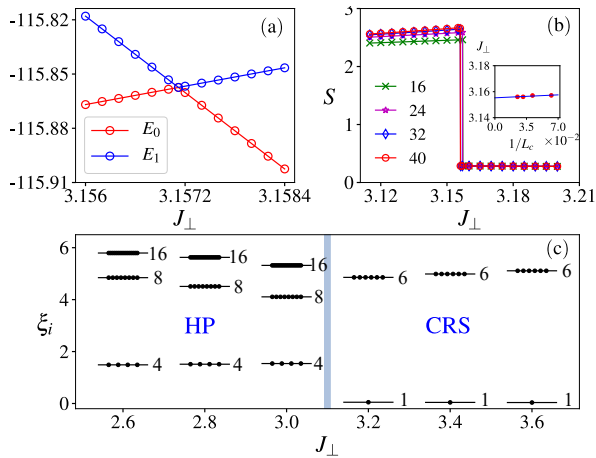}
\caption{A phase transition for $n_p=2$ is determined from $E_0$, $E_1$, 
	$S$ and $\xi_i$. Here $J_\parallel = 3$ is fixed in our calculation, and $S$, $\xi_i$
	are obtained with equal system and environment sizes.  
$\rm (a)$ $E_0$ and $E_1$ for $L_c=16$ are shown as a function of $J_\perp$. 
Level crossing occurs at $J_\perp=3.1571(2)$, indicating a first-order phase transition. 
$\rm (b)$ $S$ for the cluster number $L_c=16, 24, 32, 40$ is shown as a function of $J_\perp$. 
	The jump in $S$ indicates a first-order phase transition. 
	Inset: finite-size extrapolation to determine the transition point $J_\perp=3.156(1)$ in the thermodynamic limit. 
$\rm (c)$ Entanglement spectrum $\xi_i$ and corresponding degeneracy for $L_c=40$ in the HP phase and the CRS phase are shown. 
	In the HP phase the entanglement spectrum  
is even-fold degenerate, which is a characteristic feature of the symmetry-protected topological phase. In the CRS phase the lowest entanglement 
spectrum is nondegenerate and it is nearly zero, suggesting that the ground state is a product state of the system and the environment.}
\label{FIG3}
\end{figure}

The phase diagram and the phase boundary can be determined by the ground state and first excited state as well 
as by the entanglement entropy and entanglement spectrum.  
In the following we will illustrate our procedure that determines the phase boundary for $n_p=2$ and for other even $n_p$ it is similar.
In Fig. \ref{FIG3} (a), we plot the energy of the ground state and the first excited state for $L_c=16$
as a function of $J_\perp$. An energy-level crossing occurs at $J_\perp=3.1571(2)$, and this signals a first-order phase transition
between the HP phase and the CRS phase. Moreover, we calculate the excitation gap $\Delta$ and find that both phases are gapful. 
In Fig. \ref{FIG3}(b), we show that the entanglement entropy $S$ for $L_c=16, 24, 32$ and $40$. In this case, the length of the system and the environment is equal. 
We observe a jump in $S$. This is interpreted as a first-order phase transition. After a finite-size extrapolation, 
the transition point is determined at $J_\perp=3.156(1)$. Such phase transition is also reflected in the entanglement spectrum.
As we show in Fig.\ref{FIG3} (c), the degeneracy of the entanglement spectrum is different in the two phases. In the HP phase, all the entanglement 
spectrum is even-fold degenerate, which is a characteristic feature of the symmetry-protected topological phase\cite{PhysRevLett.96.110405,PhysRevLett.96.110404,PhysRevLett.101.010504,PhysRevB.81.064439,PhysRevLett.104.130502,PhysRevB.82.241102,PhysRevB.89.125112}. However, in the CRS phase, 
some of the entanglement spectrum is even-fold degenerate and other is odd-fold degenerate. 
\begin{figure}
\includegraphics[width=1.0\columnwidth]{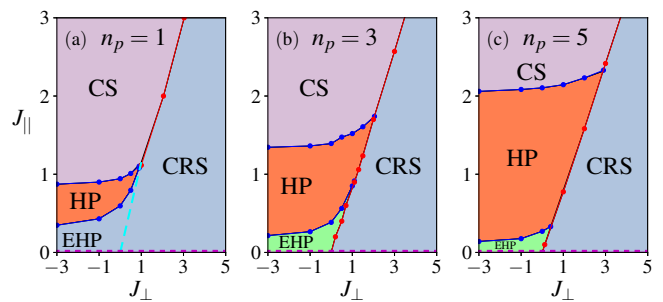}
\caption{Phase diagrams for $n_p=1,3,5$. The blue lines between the CS phase and the HP phase , and between the 
	HP phase and the EHP phase represent continuous phase transitions. The dashed cyan line in (a) 
represents a crossing between EHP phase and CRS phase instead of a phase transition. It is determined from the edge states under OBC. 
Along the dashed purple line ($J_\parallel=0$), the rungs are decoupled. Other lines denote first-order phase transitions.}\label{FIG4}
\end{figure}

\section{Phase Diagram for Odd $n_p$}
When $n_p$ is odd the phase diagrams in Fig. \ref{FIG4} are different from those for even $n_p$. 
In addition to the HP phase and the CRS phase, we find two more phases. 
One is the cluster singlet (CS) phase and the other is the EHP phase. 
Similar to the method extracting the information of the CRS phase for $n_p=2$ in the appendix \ref{appA}, 
the properties of the CS phase can be determined as well from the entanglement entropy with various cuts. 
In the CS phase, the intra-cluster interaction $J_\parallel$ is dominant and each cluster is a singlet. 
The dominant term in the ground state is the product state of these singlets. Moreover, as $J_\parallel$ increases, quantum fluctuation around such 
product state becomes smaller. Therefore, $S$ of the cuts separating one cluster converges adiabatically to a nonzero constant 
while that of the cut separating two nearest-neighbor clusters converges to zero in the large $J_\parallel$ limit.
As we explain in Sec. I, the EHP phase is different from the HP phase. Particularly, it is not protected by symmetries. 
Therefore, it can evolve into a product state accompanied with a phase transition or without a phase transition\cite{PhysRevB.81.064439}. 
$n_p=3, 5$ belong to the former case. There is a first-order phase transition between the EHP phase and the CRS phase. 
But $n_p=1$ belongs to the latter case, i.e., there is no phase transition between the EHP phase and the CRS phase,
agreeing with previous works\cite{PhysRevB.64.214413,PhysRevB.66.054435}.
Even so, these two phases can be distinguished by their edge states. To show this, we calculate low-energy states under 
both PBC and open boundary condition (OBC). 
We find that under both PBC and OBC the ground state of the CRS phase is unique. 
The ground state of the EHP phase is unique under PBC but it is nine-fold degenerate under OBC, 
demonstrating the presence of a nearly free $S=1$ effective spin at each end of the ladder. 
\begin{figure}[t]
\includegraphics[width=1.0\columnwidth]{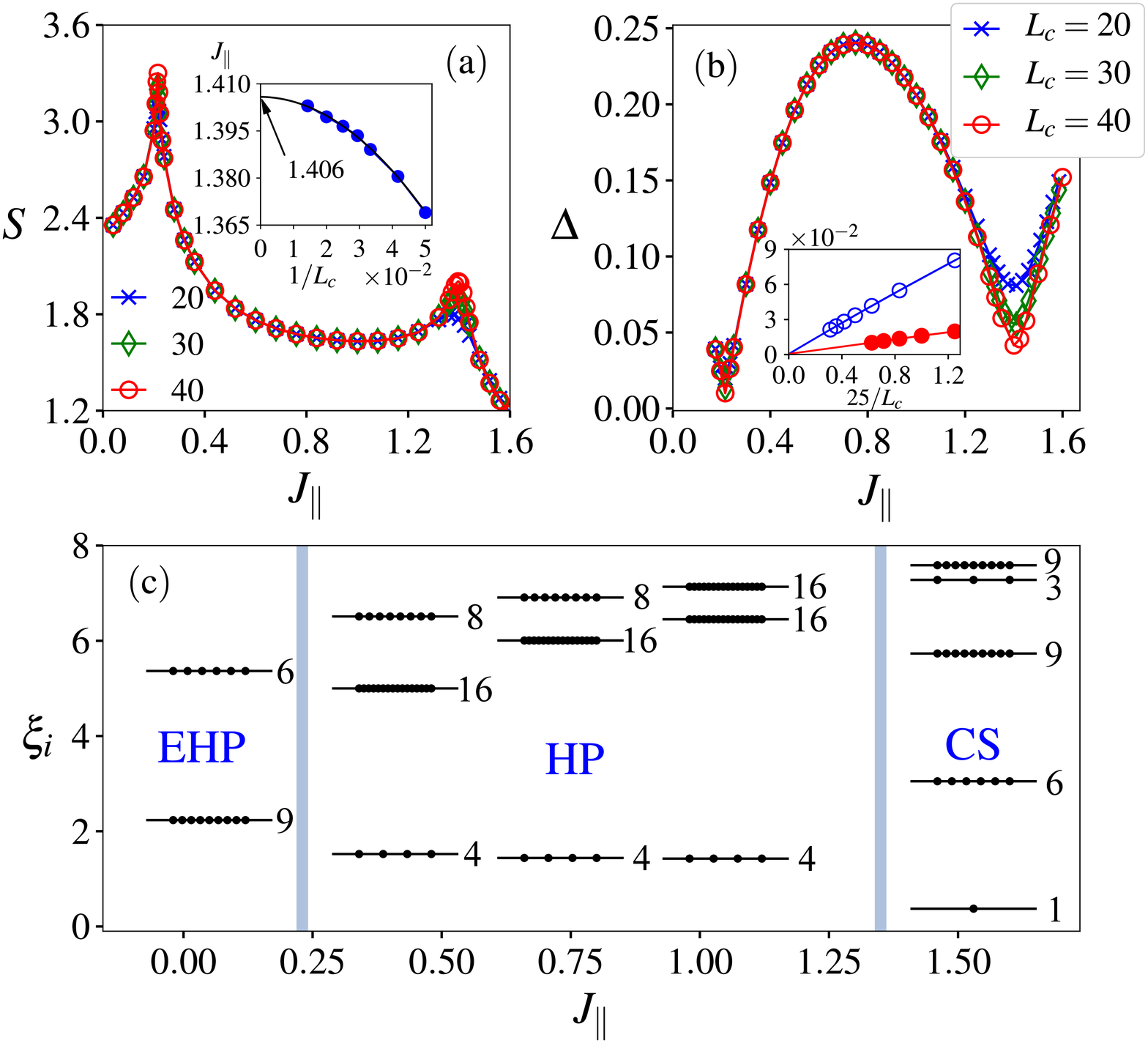}
\caption{$S$, $\Delta$ and $\xi_i$ for $n_p=3$ are plotted as a function of 
$J_\parallel$. In our calculation, $J_\perp=-3$ is fixed, and $S$, $\xi_i$ are obtained with equal system and environment sizes. 
	$\rm (a)$ $S$ for $L_c=20$, $30$ and $40$. The two peaks in $S$ suggest two phase transitions.  Inset: the positions of the right peak 
	are extrapolated to the thermodynamic limit.  
$\rm (b)$ The excitation gap for $L_c=20$, $30$ and $40$. Inset: in the thermodynamic limit, the excitation gaps at the two dips close. Red filled circles
for $L_c$ from $20$ to $40$ with a step $5$, blue open circles for $L_c$ from $20$ to $80$ with a step $10$. 
$\rm (c)$ The entanglement spectrum and corresponding degeneracy for $L_c=40$ in three phases. In the HP phase, they are even-fold degenerate.}
\label{FIG5} 
\end{figure}

\begin{figure}[t]
\includegraphics[width=1.0\columnwidth]{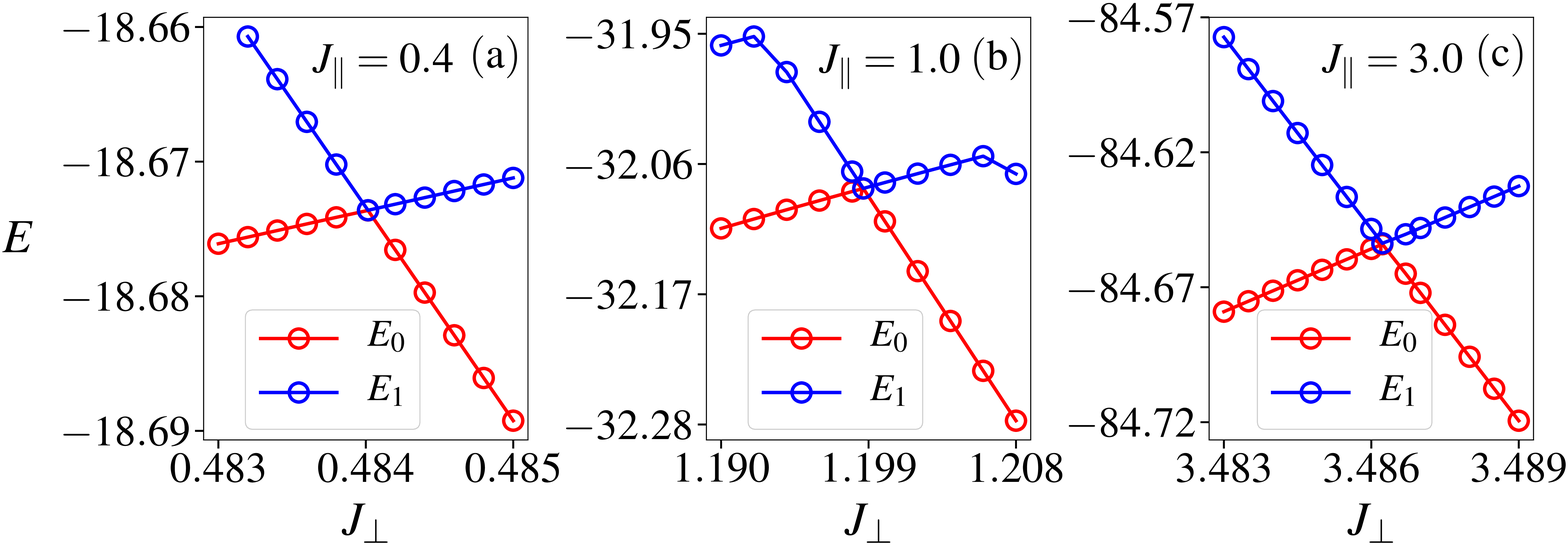}
\caption{$E_0$ and $E_1$ for $L_c=8, n_p=3$ are plotted as a function of $J_\perp$.
The level-crossing suggests a first-order phase transition.}
\label{FIG6}
\end{figure}

Similar to those with even $n_p$, the phase diagrams with odd $n_p$ are determined as well 
by $E_0$, $E_1$, $S$ and $\xi_i$. We demonstrate this for $n_p=3$ in Fig. \ref{FIG5} and Fig. \ref{FIG6}. 
Let us first see Fig. \ref{FIG5}. For simplicity, we fix $J_\perp=-3$.
In panel (a), we show the entanglement entropy $S$ for $L_c=20, 30$ and $40$ as a function of $J_\parallel$. Two sharp peaks are clearly observed, 
suggesting a phase transition near each of them. The accurate positions of the two critical points can be obtained by finite-size extrapolation, 
as we show in the inset for the right one.
In the thermodynamic limit, they are $J_\parallel=0.216(2)$ and $1.406(2)$.
Moreover, contrary to that 
in Fig. \ref{FIG3}, there is no discontinuity in $S$, and this suggests a continuous phase transition. 
In panel (b), we show the excitation gap for various $L_c$ and two minimums are found. 
In the inset, we show the extrapolation of the excitation gaps at $J_\parallel=0.216$ and $J_\parallel=1.406$. 
In the thermodynamic limit they become zero. We also confirm that the phases
are gapful. These results are in good agreement with those obtained from the entanglement entropy. 
Since a characteristic feature of the HP phase is the even-fold-degenerate entanglement spectrum, we calculate 
them with equal system length and environment length.
As we show in panel (c), the spectrum is indeed even-fold degenerate in the HP phase but in the EHP phase and CS phase they are not. 
Moreover, we calculate the degeneracy of the ground state in the HP phase under both OBC and PBC.
We find that the ground state is unique under PBC but it is 4-fold degenerate under
OBC. These provide further information supporting our phase diagrams.
In Fig. \ref{FIG6}, the two lowest energies are shown near the phase-transition points. $J_\parallel=0.4, 1.0$ and $3.0$
correspond to the EHP to CRS, HP to CRS and CS to CRS transitions, respectively. In all three cases, a level-crossing occurs, suggesting a first-order phase
transition. 

\section{Large $n_p$ Limit}
Now, we have the phase diagrams for various $n_p$. From Fig. \ref{FIG2} and \ref{FIG4}, it is obvious that these phase diagrams depend on 
the parity of $n_p$. For even $n_p$, the phase diagram has an HP phase and a CRS phase. But for odd $n_p$, 
there are four phases, an HP phase , a CRS phase , a CS phase and an EHP phase. Moreover, our results show that the region of 
the CS and EHP phases becomes smaller as $n_p$ increases. Therefore, we expect that in the large $n_p$ limit the phase diagrams will 
include only an HP phase and a CRS phase\cite{notes1}. For small $n_p$, Hamiltonian (\ref{HINTER}) can not be neglected even if $J_{\rm{inter}}$ is much 
smaller in comparison with $J_{\perp}$ or $J_{\parallel}$ because it connects the two nearest-neighbor clusters.  
However, in the large $n_p$ limit, the bulk properties are determined solely by Hamiltonian (\ref{HINTRA}).
Hamiltonian (\ref{HINTER}) only have some edge effect and thus can be neglected. 

First, we consider the properties of Hamiltonian (\ref{HINTRA}) and try to gain some insight from them.
Actually, Hamiltonian (\ref{HINTRA}) in the large $n_p$ limit has been extensively studied\cite{PhysRevB.43.8644,PhysRevB.48.10653,PhysRevB.52.12485,PhysRevB.57.11439,PhysRevB.63.144433,PhysRevB.77.214418,1367-2630-14-6-063019,PhysRevB.86.064401,PhysRevB.86.075133} and its phase diagram is already known. It includes 
an HP phase and a rung-singlet(RS) phase. A first-order phase transition is exactly known\cite{PhysRevB.52.12485} to occur at $J_\perp/J_\parallel = 1.401\cdots$.   

\begin{center}
\begin{figure}[t]
 \includegraphics[width=0.9\columnwidth]{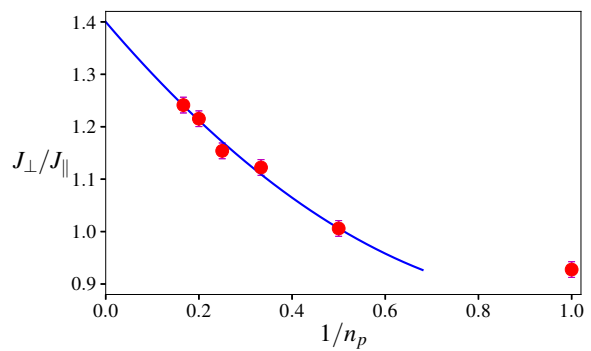}
\caption{$J_\perp/J_\parallel$, the estimated phase boundary in the large $J_\perp$ limit, is plotted as a function of  $1/n_p$.  
	The blue solid line is a quadratic polynomial fitting of the data. We have $J_\perp/J_\parallel = 1.40(3)$ as $n_p\rightarrow\infty$, which 
	is well consistent with expected result\cite{PhysRevB.52.12485} $J_\perp/J_\parallel = 1.401\cdots$.}\label{FIG7}
\end{figure}
\end{center}

To verify our analysis, let us examine the phase boundary in Fig. \ref{FIG2} and \ref{FIG4}. We notice that when $J_\perp$ is large enough 
the phase boundary is almost linear to $J_\perp$. This means that $J_\perp/J_\parallel$ is nearly a constant. In the large $n_p$ limit, we also expect that 
the phase transition depends only on $J_\perp/J_\parallel$. In Fig. \ref{FIG7}, we extrapolate $J_\perp/J_\parallel$ as a quadratic polynomial of $1/n_p$ 
to the large $n_p$ limit and obtain $J_\perp/J_\parallel=1.40(3)$, and this agrees well with the expected 
transition point $J_\perp/J_\parallel= 1.401\cdots$. 
Moreover, in the large $n_p$ limit, the edge configuration can be neglected and 
CRS phase becomes RS phase. These provide strong numerical evidence supporting our analysis.

\section{Conclusions} \label{conclusion}
In conclusion, we study cluster-based two-leg ladders with the plaquette number $n_p$ up to six.   
These models are direct extensions of the model proposed for an edge-shared tetrahedral spin-cluster compound $\rm{K_2Cu_3O\left(SO_4\right)_3}$.   
The numerically exact ground-phase phase diagrams are mapped out by using density-matrix renormalization group method.
We find that they are closely associated with the parity of $n_p$. 
For even $n_p$, there are two phases in the phase diagram, which includes an HP phase and a CRS phase. 
For odd $n_p$, in addition to the HP phase and the CRS phase, there are two more phases, which are a CS phase and an EHP phase. 
Moreover, the region of such two phases shrinks as $n_p$ increases, which leads to our conjecture that in the large $n_p$ limit they may disappear\cite{notes1}. 
By extrapolating the phase transition points to the large $n_p$ limit, we can reproduce the phase transition point of the Hamiltonian without the intercluster
coupling (i.e. $J_{\rm{inter}}=0$), which verifies our conjecture. 
By analysing the energy and entanglement entropy, we determine the order of the phase transitions. 
The transition from HP phase to CS phase or to EHP phases in odd $n_p$ are continuous. 
There is no phase transition from the EHP phase to the CRS phase for $n_p=1$, and all other phase transitions are first order. 

It was argued in Ref.~\onlinecite{PhysRevLett.120.077201} that there may be an HP phase for all even $n_p$. 
Our work show that an HP phase is present in the phase diagram for all $n_p$. This can be understood from the Hamiltonian (\ref{HINTRA}) 
in the large $n_p$ limit. In such case, there is an HP phase in the Hamiltonian (\ref{HINTRA}), no matter $n_p$ is even or odd.
This phase may persist after turning on the $J_{\rm{inter}}$ because it does not break the time-reversal symmetry.

\section*{Acknowledgments}
We acknowledge useful discussion with Zheng-Xin Liu, Gao-Yang Li, Chong Chen and Chen Cheng. This work was supported by 
NSFC(Grants No. 11874188, 11834005, 11674139, 11704166, 11474029), PCSIRT (Grant No. IRT-16R35) and 
the Fundamental Research Funds for the Central Universities.

\appendix

\section{Characteristics of the CRS phase from the entanglement entropy}\label{appA}
\begin{center}
\begin{figure}[t]
 \includegraphics[width=0.99\columnwidth,clip]{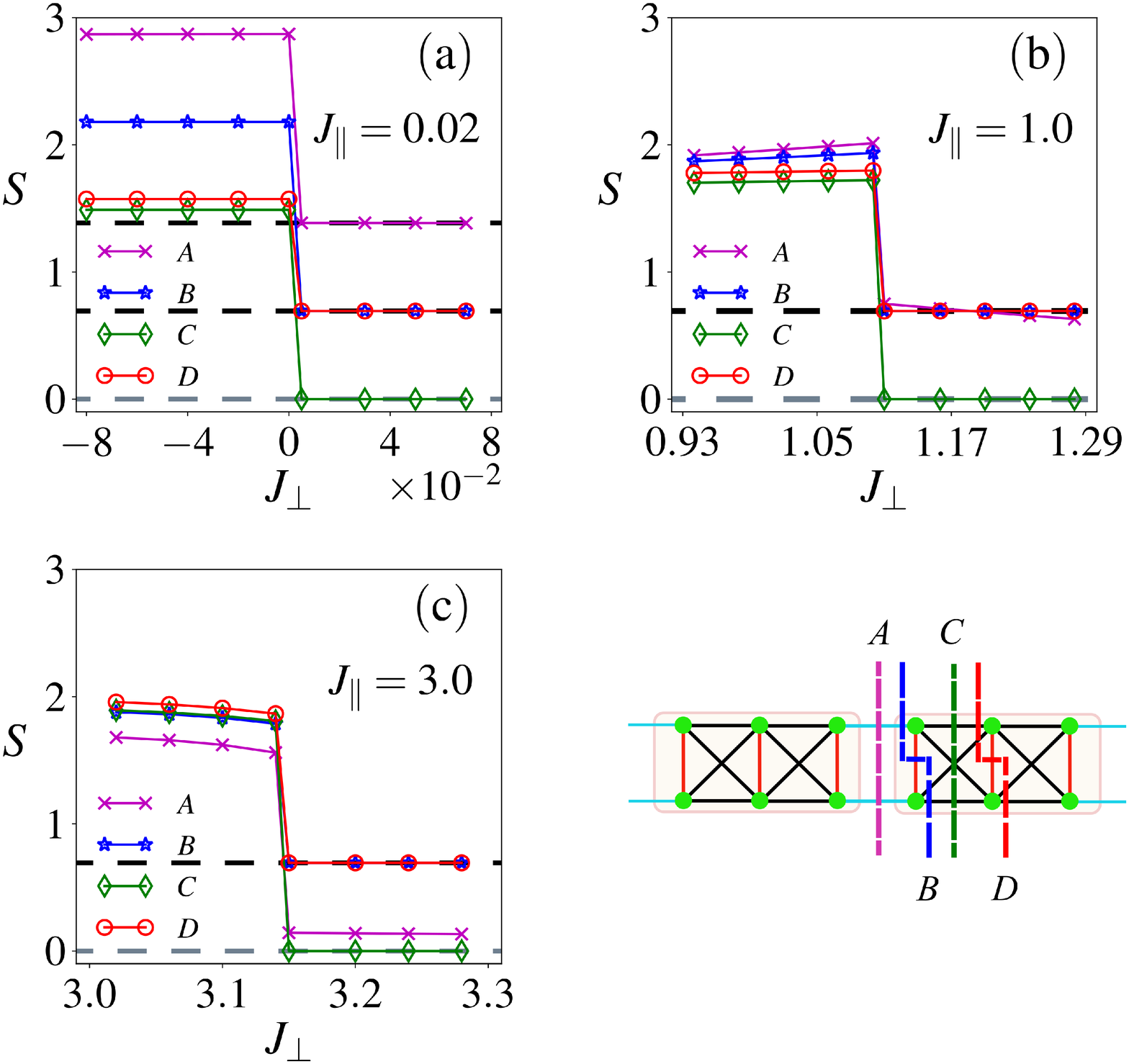}
\caption{Entanglement entropy corresponding to four different system-environment cuts $A, B, C, D$ is shown for $J_\parallel=0.02, 1.0$ and $3.0$.
        $n_p=2$, $L_c=16$, and OBC are used to illustrate our analysis. The cut $A$ separates the ladder into two equal parts. 
	The HP phase (left) and the CRS phase (right) are separated by a jump of the entanglement entropy, which signals 
	 a first-order phase transition. The black dashed line below $S=1$ is at $S=\ln2$ and that above the $S=1$ is at $2\ln2$. 
	}\label{appFIG1}
\end{figure}
\end{center}
In this Appendix, we demonstrate some properties of CRS phase in Fig. \ref{FIG2} from the 
aspect of the entanglement entropy of four different cuts. These cuts are readily available in the DMRG sweeps. 
For simplicity, we use $n_p=2$ to illustrate our analysis and for other $n_p$(both even and odd), it is similar.
Moreover, the same analysis is applicable to the CS phase.
In the following, we focus on the CRS phase only.  
As we show in Fig. \ref{appFIG1}, the entanglement entropy $S$ of the cut $B, C$ and $D$ in CRS phase is almost independent of $J_\parallel$.
In particular, $S$ of the cut $C$ is nearly zero. 
This suggests that the ground state is a product state of the system and the environment.
Moreover, $S$ of the cut $D$ is nearly $\ln2$, which is just that of a singlet formed by two spin-$\frac{1}{2}$'s. 
After considering the symmetry, we may conclude that the two spin-$\frac{1}{2}$'s connected by the rung cutted by $D$ forms a singlet. 
We notice that $S$ of the cut $B$ is nearly $\ln2$. This suggests that the rightmost two spins of the left cluster and the leftmost 
two spins of the right cluster form a singlet. However, it is not a product state of two rung singlets, 
which becomes clear when we study the $S$ of the cut $A$. To reflect such difference from the RS phase, we use CRS to mark this phase.

The nearly zero $S$ of the cut $C$ suggests that the largest $\rho_i$ is almost $1$ and all others are nearly zero. In the DMRG simulations, 
it is thus difficult to select optimal bases according to $\rho_i$. To obtain accurate results, more bases should be kept, in particular, for excited states.


\begin{thebibliography}{58}
\bibitem [{E11()}]{notes1} Strickly speaking, this conclusion is valid for any finite $J_\parallel$ and $J_\perp$.
\end{thebibliography}
%
\end{document}